\documentclass[12pt]{iopart}
\usepackage{graphicx}

\newcommand{\gtrsim}{\raisebox{-0.13cm}{~\shortstack{$>$ \\[-0.07cm] $\sim$}}~}
\newcommand{\lesssim}{\raisebox{-0.13cm}{~\shortstack{$<$ \\[-0.07cm] $\sim$}}~}

\begin{document}

\title[DQPTs in the one-dimensional extended Fermi-Hubbard model]{Dynamical quantum phase transitions in the one-dimensional extended Fermi-Hubbard model}

\author{Juan Jos\'e Mendoza-Arenas}

\address{Departamento de F\'{i}sica, Universidad de los Andes, A.A. 4976, Bogot\'a D. C., Colombia}
\address{H.H. Wills Physics Laboratory, University of Bristol, Bristol BS8 1TL, UK}

\ead{jj.mendoza@uniandes.edu.co}

\begin{abstract}
We study the emergence of dynamical quantum phase transitions (DQPTs) in a half-filled one-dimensional lattice described by the extended Fermi-Hubbard model, based on tensor network simulations. Considering different initial states, namely noninteracting, metallic, insulating spin and charge density waves, we identify several types of sudden interaction quenches which lead to DQPTs. Furthermore, clear connections to particular properties of observables, specifically the mean double occupation or charge imbalance, are established in two main regimes, and scenarios in which such correspondence is degraded and lost are discussed. Dynamical transitions resulting solely from high-frequency time-periodic modulation are also found, which are well described by a Floquet effective Hamiltonian. State-of-the-art cold-atom quantum simulators constitute ideal platforms to implement several reported DQPTs experimentally.
\end{abstract}

% Uncomment for keywords
\vspace{2pc}

% Uncomment if a separate title page is required
\maketitle

\section{Introduction}
In spite of being the object of intense research for several decades, interacting many-body quantum systems continue posing some of the most exciting and challenging problems in modern physics. Key examples manifest during their unitary dynamics resulting from nonequilibrium setups such as sudden quenches, where fundamental phenomenology including thermalization \cite{rigol2008nat,nandkishore2015annu,eisert2015nat,kaufman2016science,ueda2020nat} and transport of conserved quantities \cite{schneider2012nat,jepsen2020nat,bertini2021rmp} emerges. The understanding of these problems has received an enormous boost largely due to the development of powerful numerical methods \cite{schollwock2011ann,eckstein2014rmp,orus2019nat}, and due to the implementation of quantum simulators which allow for exquisite unprecedented control of the degrees of freedom of many-body systems \cite{georgescu2014rmp,schafer2020nat,altman2021prx}.

A fascinating effect identified within this combined effort, known as dynamical quantum phase transitions (DQPT), results from taking general ideas of quantum criticality to the nonequilibrium scenario~\cite{heyl2013prl,heyl2018rev}. This concept relies on the return (or Loschmidt) amplitude of the evolved state $|\psi_0(t)\rangle$ of a quantum system to its initial state $|\psi_0\rangle$,
\begin{eqnarray}
\mathcal{G}(t)=\langle\psi_0|\psi_0(t)\rangle=\langle\psi_0|e^{-iHt}|\psi_0\rangle,
\end{eqnarray}
during the dynamics arising from a sudden quench. Here $|\psi_0\rangle$ is the ground state of a Hamiltonian $H_0$, and $H$ is the post-quench Hamiltonian. The Loschmidt amplitudes have been argued to be similar to canonical partition functions in equilibrium~\cite{heyl2013prl}. Namely, for a system of $L\gg1$ sites, the associated Loschmidt echo $\mathcal{L}(t)$ has a dependence with $L$ of the form
\begin{eqnarray}\label{echo}
\mathcal{L}(t)=|\mathcal{G}(t)|^2=e^{-L\lambda(t)}
\end{eqnarray}
with $\lambda(t)$ a real-valued (intensive) rate function which is obtained in the thermodynamic limit as
\begin{eqnarray} \label{rate_1}
\lambda(t)=-\lim_{L\to\infty}\frac{1}{L}\log\left[\mathcal{L}(t)\right].
\end{eqnarray}
A vanishing value of the return amplitude, and thus of the Loschmidt echo, results in nonanalyticities of the rate function $\lambda(t)$. This is similar to how the complex zeros of partition functions, known as Fisher or Lee-Yang zeros, lead a nonanalytic behavior of the free energy density (i.e. the associated rate function) at the critical point of a phase transition.
%such as the critical temperature $T_c$ for temperature-driven transitions.
Extending this idea to the nonequilibrium realm, a DQPT at a critical time $t^{*}$ (instead of a critical control parameter) is said to occur when during the dynamics, such zeros are crossed~\cite{heyl2013prl}.
  
A vast amount of recent theoretical research has been devoted to study this beautiful insight in systems of different nature, namely spin~\cite{heyl2013prl,karrasch2013prb,heyl2014prl,andraschko2014prb,szabolcs2014prb,heyl2015prl,markus2015prb,divakaran2016pre,stauber2017prb,homrighausen2017prb,zunkovic2018prl,kennes2018prb,sun2020prb,haldar2020prb,halimeh2020prr,ding2020prb,nicola2021prl,gonzalez2022}, fermionic~\cite{canovi2014prl,jafari2019prb,pastori2020prr,uhrich2020prb,rylands2020,rylands2021}, bosonic~\cite{fogarty2017njp,mateusz2019prb,abdi2019prb,liao2020,syed2021prb,stumper2021arxiv}, and hybrid models~\cite{puebla2020prb}. This effort also includes the analysis of the impact of ingredients such as disorder~\cite{yin2018pra,cao2020prb,mishra2020jpa} and topological order~\cite{budich2017prb,hagymasi2019prl,sadrzadeh2021prb}. Crucially, experimental demonstrations of DQPTs have been achieved in several quantum simulation platforms~\cite{jurcevic2017prl,flaschner2018nat,guo2019prap,wang2019prl,nie2020prl}. Many of these studies have shown evidence that DQPTs emerge when there is a quench across an equilibrium quantum phase transition, i.e. when $|\psi_0\rangle$ and the ground state of $H$ correspond to different phases. However, several exceptions have been reported~\cite{heyl2018rev,andraschko2014prb,szabolcs2014prb,stumper2021arxiv,jafari2019sci}, so no one-to-one correspondence between DQPTs and equilibrium quantum phase transitions can be established. In addition, it is common to look for manifestations of DQPTs beyond nonanalyticities of $\lambda(t)$. In several scenarios, such as particular quenches in the XXZ~\cite{heyl2014prl,hagymasi2019prl}, O$(N)$~\cite{weidinger2017prb} and symmetry-broken Ising~\cite{jurcevic2017prl} models, these nonanalyticities coincide with vanishing values of order parameters; in other cases they are shifted from zeros or minima of order parameters by a constant factor, as found in Ising~\cite{heyl2013prl,zunkovic2018prl} and Bose-Hubbard models~\cite{mateusz2019prb} respectively. However, in many others such an association remains elusive~\cite{heyl2013prl,heyl2018rev,karrasch2013prb,heyl2014prl}.

To help unravel the connections between DQPTs, equilibrium quantum criticality and the time evolution of observables, it is valuable to consider simple models with different phases that are accessible experimentally. In this regard, systems of strongly interacting spin-$1/2$ fermions constitute very attractive candidates to be the object of such analysis. In spite of this, DQPTs on these systems have been discussed only a few times ~\cite{canovi2014prl}. In the present work we perform such a study in the one-dimensional extended Fermi-Hubbard (EFH) model at half filling, which incorporates on-site and nearest-neighbor density-density interaction. This model not only generalizes the seminal and widely-studied Fermi-Hubbard model, whose dynamics has been intensively explored experimentally with cold atoms in optical lattices \cite{schneider2012nat,ronzheimer2013prl,schreiber2015sci,scherg2018prl,tarruell2018}; it is also known to possess a rich ground-state phase diagram \cite{torsten1999prb,nakamura2000prb,deng2006prb,iemini2015prb}, and has been recently proposed to explain properties of one-dimensional quantum materials \cite{chen2021,wang2021}. We analyze DQPTs emerging at early times from quantum quenches in one or both of the interaction terms, with special focus on the Fermi-Hubbard limit and charge density wave (CDW) states. For the former case, we also discuss how symmetries of the model are manifested in the time evolution of the rate function. Furthermore, we establish two main general scenarios where a clear connection between DQPTs and the dynamics of observables exists. Finally we show that DQPTs can be induced solely by a time periodic modulation of an on-site potential, which provides a timely strategy for observing such phenomena given the state-of-the-art advances in Floquet engineering with cold fermions in optical lattices \cite{gorg2018nat,messer2018prl,sandholzer2019prl,viebahn2021prx}.

The manuscript is organized as follows. In section \ref{model_sect} we describe the EFH model and the method used to study its dynamics. In section \ref{dqpt_fh} we show the existence of DQPTs in the Fermi-Hubbard limit ($V=0$) for noninteracting, weakly and strongly interacting initial states. The transitions resulting from CDW states, namely a ground state of the extended model for finite $V>0$ and a product degenerate CDW, are discussed in section \ref{dqpt_cdw}. The emergence of DQPTs from Floquet modulation is presented in section \ref{floquet}. Finally, section \ref{conclu} contains our main conclusions.

\section{Model and method} \label{model_sect}

We study the DQPTs of the one-dimensional EFH model at half filling, zero magnetization and with open boundary conditions. Its Hamiltonian is given by
\begin{eqnarray}\label{EFH_1D}
\hat{H}=-J\sum_{j=1}^{L-1}\sum_{\sigma=\uparrow,\downarrow}(\hat{c}_{j,\sigma}^{\dagger}\hat{c}_{j+1,\sigma}+{\rm H.c.})+U\sum_{j=1}^L\hat{n}_{j\uparrow}\hat{n}_{j\downarrow}+V\sum_{j=1}^{L-1}\hat{n}_j\hat{n}_{j+1},
\end{eqnarray}
where $\hat{c}_{j,\sigma}^{\dagger}$ ($\hat{c}_{j,\sigma}$) creates (annihilates) a fermion with spin $\sigma=\uparrow,\downarrow$ on site $j$, $\hat{n}_{j\sigma}=\hat{c}_{j,\sigma}^{\dagger}\hat{c}_{j,\sigma}$ is the number operator for site $j$ and spin $\sigma$, $\hat{n}_j=\hat{n}_{j\uparrow}+\hat{n}_{j\downarrow}$ is the total number operator at site $j$, $J$ is the hopping (taken as $J=1$ to set the energy scale), $U$ the on-site coupling, and $V$ the nearest-neighbor interaction. Thus this model extends the standard (integrable) Fermi-Hubbard Hamiltonian, which corresponds to $V=0$. Importantly, we allow $U$ and $V$ to represent both repulsive ($U,V>0$) or attractive ($U,V<0$) interactions.

\begin{figure}[t]
\begin{center}
\includegraphics[scale=1.15]{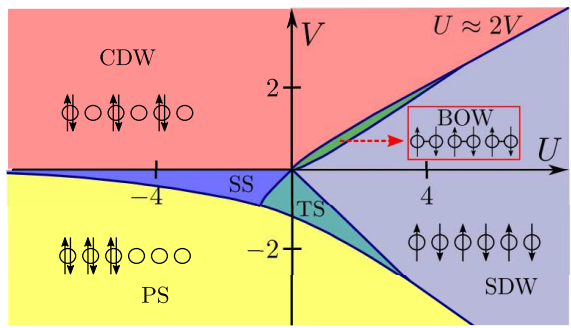}
\end{center}
\caption{Qualitative depiction of the ground-state phase diagram of the one-dimensional EFH model at half filling and zero magnetization \cite{iemini2015prb}.}
\label{fig1}
\end{figure}

The ground state phase diagram of the model at half filling and zero magnetization, i.e. with
\begin{eqnarray} \label{symm}
\sum_{j=1}^L\langle \hat{n}_{j\uparrow}\rangle=\sum_{j=1}^L\langle \hat{n}_{j\downarrow}\rangle=\frac{L}{2},
\end{eqnarray}
is well known~\cite{torsten1999prb,nakamura2000prb,deng2006prb,iemini2015prb}, and features several (quasi-long range ordered) phases, as sketched in figure \ref{fig1}. 
%These are separated by different first, second or infinite (Berezinskii-Kosterlitz-Thouless) order quantum phase transitions. 
When the system has on-site attraction and nearest-neighbor repulsion, or when the nearest-neighbor repulsion dominates over on-site repulsion, a charge density wave (CDW) is favored. On the other hand, if the on-site repulsion is dominant, a spin density wave (SDW) is formed. For weak and intermediate repulsive interactions, a bond-order wave (BOW) emerges between the CDW and SDW states, characterized by an alternating expectation value of the hopping term of the Hamiltonian~\cite{nakamura2000prb,sengupta2002prb,jeckelmann2002prl,zhang2004prl,sandvik2004prl,ejima2007prl,dalmonte2015prb,spalding2019prb}. For strong nearest-neighbor attraction, the fermions cluster together forming a phase separation (PS). Finally, for weak and intermediate nearest-neighbor attraction, superconducting states are established, either of singlet (SS, with on-site attraction) of triplet (TS) pairing \cite{nakamura2000prb,lin2000chin}.  

The dynamical properties of the one-dimensional EFH model are known to a much lesser degree, since most research has focused on the $V=0$ limit. The latter includes analysis on the expansion of initially-confined particles~\cite{heidrich-meisner2008pra,heidrich-meisner2009pra,karlsson2011epl,kajala2011prl,kessler2012pra,langer2012pra,chien2012pra,kessler2013njp,vidmar2013prb,lacroix2014prb,mei2016pra,schlunzen2017prb,scherg2018prl,cook2020pra}, doublon dynamics~\cite{rausch2017prb}, melting of product ordered states~\cite{enss2012njp,vidmar2013prb,heidrich_meisner2015pra,schlunzen2017prb}, and quantum quenches from ground~\cite{vidmar2013prb,hamerla2013prb,ido2015prb,bleicker2018pra,bleicker2020pra} and thermal states~\cite{goth2012prb,white2019prb}. On the other hand, for finite nearest-neighbor interactions, the study on dynamical properties has been much more limited, focusing on the time evolution of doublons~\cite{hofmann2012prb}, time-resolved single-particle spectrum~\cite{shao2020prb}, and spectral functions for detecting nonequilibrium superconductivity \cite{paeckel2020prb}. Thus the nonequilibrium properties of the EFH model still constitute a largely uncharted territory.

In the present work we discuss the time evolution induced by different sudden quenches in the EFH model, focusing on early times where the existence of DQPTs is unveiled. We consider three main scenarios. First we discuss transitions solely within the Fermi-Hubbard limit. Then we explore quenches from CDW states, corresponding to the ground state of the EFH with finite $V$ or a product state. Finally we study DQPTs emerging from a periodic modulation of the Hamiltonian. To reach systems of hundreds of sites, our simulations are based on a matrix product state description. Namely, we calculate the initial state $|\psi_0\rangle$ as the ground state of a Hamiltonian $H_0$ (with interactions $U_0$ and $V_0$) with the density matrix renormalization group~\cite{schollwock2011ann}, and use its time-dependent extension~\cite{vidal2004prl,paeckel2019ann} to obtain the evolved state $|\psi_0(t)\rangle$ under Hamiltonian $H$ (with interactions $U$ and $V$).  We show results for chains of up to $L=160$ for correlated initial states, and of $L=256$ for the product CDW (finite-size effects are discussed in \ref{appendixA}). We also incorporate the conservation of number of fermions with spin up and down, to work directly in the symmetry sector of half filling and zero magnetization \eref{symm}. The ground states have been calculated with maximal truncation error per sweep down to $10^{-10}$. In addition, during the dynamics we take time steps down to $\delta t=0.001$ and maximal truncation errors per step down to $10^{-10}$ (up to a maximal truncation parameter $\chi=3000$). Our codes are based on the open-source Tensor Network Theory library \cite{tnt,tnt_review1}.

\section{DQPT in the Fermi-Hubbard model} \label{dqpt_fh}

We initiate by discussing DQPTs in the Fermi-Hubbard limit $V_0=V=0$. In this case, in addition to the rate function $\lambda(t)$ of ~\eref{rate_1}, we calculate the average double occupation in the lattice, given by
\begin{eqnarray} \label{mean_d}
d(t)=\frac{1}{L}\sum_{j=1}^L\langle\hat{n}_{j\uparrow}\hat{n}_{j\downarrow}\rangle(t).
\end{eqnarray}
We also consider two distinct cases: when the initial state $|\psi_0\rangle$ is noninteracting ($U_0=0$) and when it has finite on-site interactions ($U_0\neq0$).

\subsection{Initial noninteracting state} \label{dqpt_fh_noint}

\begin{figure}[t]
\begin{center}
\includegraphics[scale=0.75]{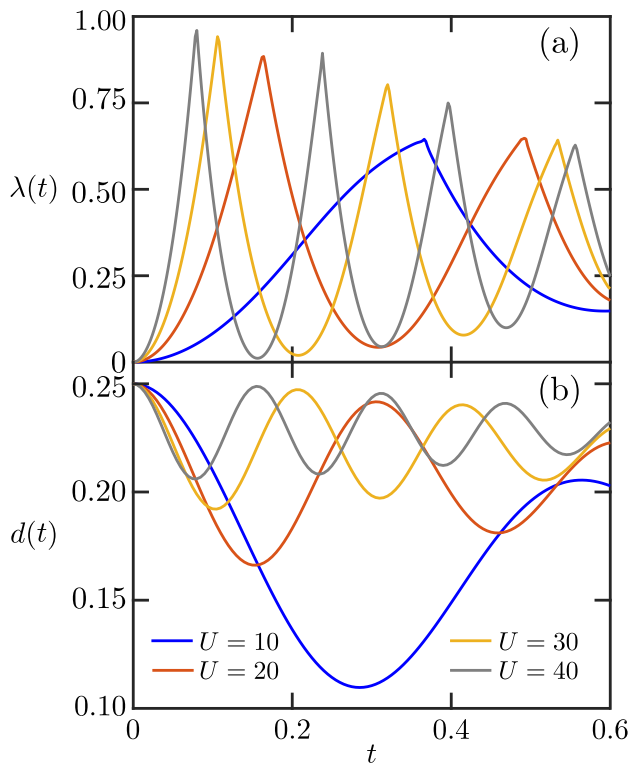}
\end{center}
\caption{(a) Rate function $\lambda(t)$ as a function of time for quenches of the Fermi-Hubbard model, with an initial state of $U_0=0$. (b) Time evolution of double occupation. The second maximum of $\lambda(t)$ and minimum of $d(t)$ for $U=10$, not shown for clarity, take place at $t_2^*=1.18$ and $t_{2}^{\rm d}=0.81$ respectively.}
\label{Rate_Hubbard_V0_Ugs0}
\end{figure}

We first describe the scenario with $U_0=0$ and $U\gg1$. Here we expect short-time dynamics governed by oscillations of frequency $U$, as already seen in small systems. For instance, using strong-coupling time-dependent perturbation theory (see \ref{appendix0}), the Loschmidt echo for two sites has the form
\begin{eqnarray}
\mathcal{L}(t)=\cos^2\left(\frac{Ut}{2}\right)+\frac{16}{U^2}\sin^2\left(\frac{Ut}{2}\right).
\end{eqnarray}
Dynamical quantum phase transitions are evidenced with the same periodicity in much larger systems; this is shown in figure~\ref{Rate_Hubbard_V0_Ugs0}(a) for $L=128$. The cusps featured by the rate function at certain times $t_j^*$ manifest the emergence of DQPTs for $U\gg1$, with earlier and more frequent transitions when $U$ increases. In fact, we have verified that the critical time $t_1^*$ of the first DQPT decays as $\sim 1/U$. Also, the differences $\Delta t_j^*=t_{j+1}^*-t_j^*$ between consecutive critical times $t_j^*$ of the transitions shown in figure~\ref{Rate_Hubbard_V0_Ugs0}(a) are the same for each value of $U$, and also decay as $\sim 1/U$.

The evolution of the corresponding double occupations is shown in figure~\ref{Rate_Hubbard_V0_Ugs0}(b), which for short times follows the same form already seen in small systems (see \ref{appendix0}), namely
\begin{eqnarray} \label{dbl_1stcase}
d_0-d(t)\propto\frac{1}{U}\sin^2\left(\frac{Ut}{2}\right),
\end{eqnarray}
with $d_0=1/4$ the mean double occupation of the noninteracting state at zero magnetization and half filling. Even though the double occupation does not constitute an order parameter of the initial state, the relation between its time evolution and DQPTs is apparent. For values of $U\gtrsim20$ the agreement between $t_j^*$ and the times $t_j^{\rm d}$ of minimal double occupation, given by
\begin{eqnarray}
t_j^{\rm d}=\frac{\pi}{U}\left(2j-1\right)\quad {\rm with}\quad j=1,2,...
\end{eqnarray}
according to \eref{dbl_1stcase}, is excellent; both times are almost identical for $U=40$ ($t_{1,2,3}^*=0.079,0.238,0.396$, $t_{1,2,3}^{\rm d}=0.078,0.235,0.395$). These results indicate that in this regime of large $U$ and early times, both the DQPTs and charge dynamics are characterized by the same timescale, given by $U$. For much lower (yet still strong) interactions this correspondence is lost. For example, for $U=10$ the times $t_j^*$ and $t_j^{\rm d}$ no longer agree, as seen in figure 2(a); the same occurs for their differences $\Delta t_j^*$ and $\Delta t_j^{\rm d}=t_{j+1}^{\rm d}-t_j^{\rm d}$. Thus the dynamics becomes more involved as both quantities are no longer governed by the same underlying nonequilibrium timescale. 

We note that for low values of $U$ we did not find DQPTs on early times; for example, for $U=2$, no DQPT was seen up to the reached time $t=1.4$. However this does not preclude their appearance much later, which might happen even when the rate function remains smooth for a long time \cite{karrasch2013prb}. 

%For lower $U$ there are also DQPT; the lowest for which I have observed considering the times reached is $U=6$ for $t=0.78$; there no agreement with minimum of $d(t)$ is seen.

\subsection{Initial interacting state} \label{ini_inter_hubbard}

Now we discuss the emergence of DQPTs when crossing the metal-insulator transition of the Fermi-Hubbard model at $U=0$. For this, we prepare initial states $|\psi_0\rangle$ for different finite interactions $U_0$ and perform a quench with values of $U$ at the other side of the transition (no DQPTs were observed for $U$ within the same equilibrium phase). We particularly focus on how cusps of the rate function evolve as $U$ is increased.

\begin{figure}[t]
\begin{center}
\includegraphics[scale=0.75]{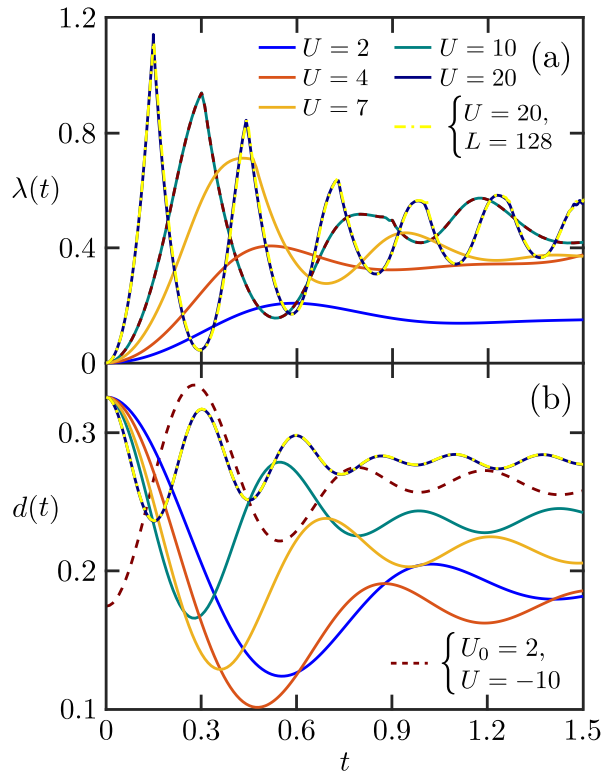}
\end{center}
\caption{(a) Rate function $\lambda(t)$ as a function of time for quenches of the Fermi-Hubbard model, with an initial state of $U_0=-2$. (b) Time evolution of double occupation.}
\label{Rate_Hubbard_V0_Ugs-2}
\end{figure}

We first take $U_0=-2$, which corresponds to a metallic state; the rate function for $L=160$ is shown in figure~\ref{Rate_Hubbard_V0_Ugs-2}(a). For weak repulsive interactions $U\leq4$, $\lambda(t)$ remains smooth; however a small maximum is featured, which grows and moves to earlier times as $U$ increases. For stronger coupling, e.g. $U=7$, the first peak starts developing a sharp behavior and more smooth peaks arise. This behavior anticipates the emergence of DQPTs for even stronger interactions, namely $U\geq10$, whose rate function is very similar to that of the noninteracting case as intuitively expected. In particular, the critical time of the first cusp $t_1^*$ also follows a $1/U$ trend, with slightly lower values than those of $U_0=0$. For longer times a few periodic DQPTs are still observed for very strong interactions $U=20$, followed by a sequence of smooth low peaks. 

The corresponding double occupations are shown in figure~\ref{Rate_Hubbard_V0_Ugs-2}(b). In general, as $U$ increases more oscillations emerge, with decreasing values of the times $t_j^{\rm d}$ of its minima. For very strong interactions $U=20$ and early times the double occupation evolves following a trend similar to that in equation \eref{dbl_1stcase}, and its first three minima agree with the nonanalyticities of the rate function. This is accentuated even further for stronger interactions, namely $U=30$ (not shown), where five DQPTs coincident with minima of the double occupation are observed. Thus, similarly to the $U_0=0$ case, in the very-strongly-interacting limit of the quench Hamiltonian there is a transparent connection between DQPTs and observable dynamics.

\begin{figure}[t]
\begin{center}
\includegraphics[scale=0.75]{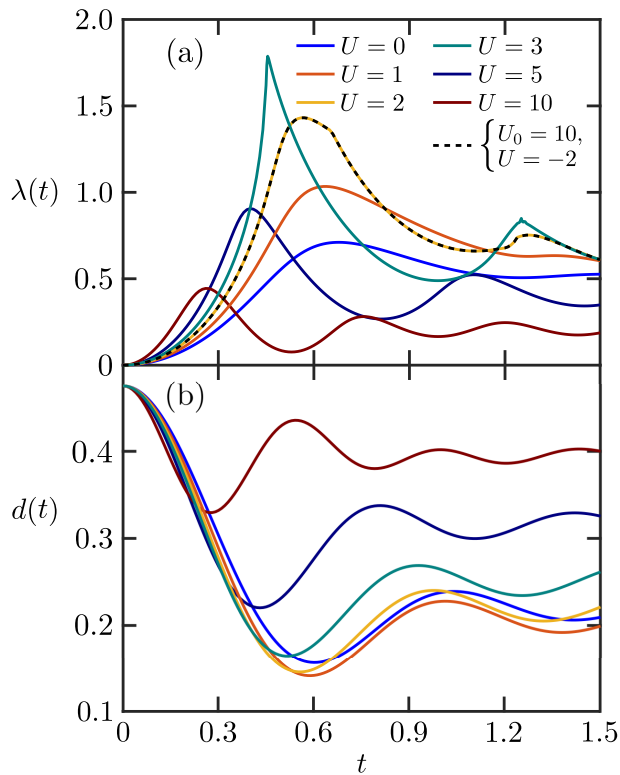}
\end{center}
\caption{(a) Rate function $\lambda(t)$ as a function of time for quenches of the Fermi-Hubbard model, with an initial state of $U_0=-10$. (b) Time evolution of double occupation.}
\label{Rate_Hubbard_V0_Ugs-10}
\end{figure}

An important feature of the time evolution of the Fermi-Hubbard model is evidenced in figure~\ref{Rate_Hubbard_V0_Ugs-2}(a) with a particular example. Namely, the rate function is identical to that obtained when the signs of $U_0$ and $U$ are changed (dashed lines). In addition, $d(t)$ is symmetric around the value $1/4$ for this modified time evolution, as depicted in figure~\ref{Rate_Hubbard_V0_Ugs-2}(b). Thus, when the initial state is prepared in the insulating phase and the time evolution is performed under a Hamiltonian with metallic ground state, the correspondence between DQPTs and the mean double occupation takes place for local maxima of the latter. We show in \ref{appendixB} how these properties arise from the symmetries of the Hamiltonian.

Now we consider an initial state of $U_0=-10$, also in the metallic phase, for $L=128$. In contrast to the previous cases, DQPTs are seen to emerge at intermediate interactions only, thus being a more restricted scenario for their implementation. Specifically, as observed in figure~\ref{Rate_Hubbard_V0_Ugs-10}(a), no evidence of DQPTs is found for $U=0,1$, and at $U=2$ sharp features close the maxima of the rate function start developing, anticipating DQPTs. The latter occur at $U=3$, and disappear again for stronger interactions $U\geq5$. The symmetry under sign inversion is also shown, exemplified for $U=2$. In figure~\ref{Rate_Hubbard_V0_Ugs-10}(b) we depict the corresponding time evolution of the double occupation. Similarly to the maxima of $\lambda(t)$, the minima of $d(t)$ emerge earlier as $U$ increases, occuring at similar (yet not always equal) times. Namely, the DQPTs of $U=3$ take place at $t_{1,2}^*=0.455,1.253$, and the $d(t)$ minima at $t_{1,2}^{\rm d}=0.510,1.255$. The lack of a full coincidence of $t_j^*$ and $t_j^{\rm d}$ or of their difference $\Delta t_j^*$ and $\Delta t_j^{\rm d}$ prevents establishing a definitive link between DQPTs and observable dynamics for this case. It is tempting to propose that interactions in a similar scenario (i.e. large $U_0$, low or intermediate $U$) can be finely tuned to induce such agreement. Whether this is the case remains an open question.

Finally, it is worth noting that from the reported results for the reached times, an asymmetry in the directions of the quenches exists. For example, taking $U_0=0$ and $U=10$ induces a DQPT (figure~\ref{Rate_Hubbard_V0_Ugs0}) at early times, while considering $U_0=10$ and $U=0$ (figure~\ref{Rate_Hubbard_V0_Ugs-10}) does not.

\section{DQPT from CDW states} \label{dqpt_cdw}
Now we discuss DQPTs from different initial CDW states, namely the ground states of the EFH model for finite (section \ref{dqpt_finite_cdw}) and infinite (section \ref{dqpt_product_cdw}) nearest-neighbor attraction $V$. In addition to $d(t)$, here we calculate the mean imbalance between the population of odd and even sites, given by
\begin{eqnarray}\label{mean_I}
I(t)=\frac{1}{L}\sum_{j=1}^L(-1)^j\langle\hat{n}_j\rangle.
\end{eqnarray}
In contrast to cases discussed in section \ref{dqpt_fh}, here the imbalance shows a sizeable nontrivial dynamical behavior.

\subsection{Finite-coupling CDW} \label{dqpt_finite_cdw}

\begin{figure}[h]
\begin{center}
\includegraphics[scale=0.75]{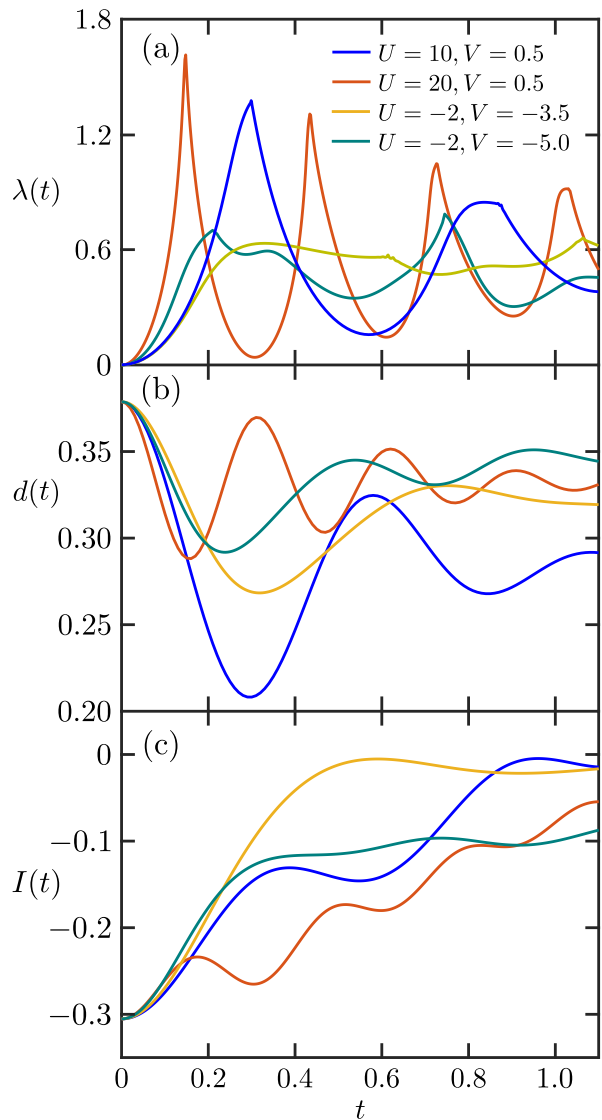}
\end{center}
\caption{(a) Rate function $\lambda(t)$ as a function of time for quenches of the Fermi-Hubbard model, with an initial state of $U_0=-2$, $V_0=0.5$. (b) Time evolution of double occupation. (c) Time evolution of population imbalance. The average of the latter was performed slightly modifying \ref{mean_I}, i.e. we removed 10 sites from each boundary of the chain. In contrast to $d(t)$, the value of the imbalance for this initial state strongly depends on the number of sites removed from the boundaries, although the times of extremal values do not. The curves for $U=20,V=0.5$ and $U=-2,V=-5$ correspond to $L=128$; those of $U=10,V=0.5$ and $U=-2,V=-3.5$, which feature a more complex behavior, are for $L=160$.}
\label{CDW_U-2V05}
\end{figure}

We start by considering DQPTs when the integrability of the Fermi-Hubbard model is broken by including nearest-neighbor interactions $V_0,V\neq0$. In particular, we consider an initial state with parameters $U_0=-2$ and $V_0=0.5$, located in the CDW phase and close to the boundary to the SS (see figure \ref{fig1}). The results are shown in figure \ref{CDW_U-2V05}.

Initially we fixed $V=V_0$ and took several values of $U>0$, corresponding to quenches into the BOW and SDW states. Similarly to the results in figures \ref{Rate_Hubbard_V0_Ugs0} and \ref{Rate_Hubbard_V0_Ugs-2}, for the times reached in the simulations we only found DQPTs for large values of $U$, deep into the SDW phase. The nonanalyticities of the first few peaks are also similar to the ones reported in those cases. For very strong on-site interaction $U=20$, these are close to the times of the local minima of $d(t)$ ($t_{1,2}^*=0.148,0.434$, $t_{1,2}^{\rm d}=0.155,0.470$), although not as close as in the $V=0$ case of figure \ref{Rate_Hubbard_V0_Ugs0}. Thus, introducing a nearest-neighbour interaction $V$ tends to blur the coincidence of DQPTs and observable dynamics, which nevertheless remains as long as $U$ is largely dominant. This is also seen from the evolution of the imbalance, which oscillates with frequency $\sim U$ and thus features maxima very close to the minima of the double occupation.

Second, we study a sequence of quenches in a vertical line of the phase diagram of figure \ref{fig1}, fixing $U=-2$ and varying $V$ so evolution Hamiltonians with SS and PS ground states are considered. We only found DQPTs after crossing to the PS regime, and show some examples in figure~\ref{CDW_U-2V05}. For $V=-3.5$, so the interaction scales are similar, only one DQPT at $t=1.06$ is clear; the nonanalytic features around $t=0.6$ decrease with system size (as verified from $L=128$ results), suggesting a finite-size effect. Moreover, no connection to observable dynamics is found. On the other hand, when $V=-5$ so one interaction parameter is dominant over the other energy scales, the DQPTs are close to the local minima of $d(t)$ but do not coincide ($t_{1,2}^*=0.210,0.745$, $t_{1,2}^{\rm d}=0.240,0.720$). However, it is natural to anticipate that for increasing values of $V$, an agreement will be established similarly to the effect of largely incrementing $U$.

Finally, we performed quenches varying $U$ and $V$, to BOW \cite{ejima2007prl,spalding2019prb} ($U=2$ and $V=1$, $U=4$ and $V=2$) and TS ($U=0$ and $V=-0.3,-0.5,-0.8$) regimes. For the reached times $t\approx2$, no DQPTs were observed (not shown). 

\subsection{Degenerate product CDW} \label{dqpt_product_cdw}

Here we analyze the dynamics of an initial product CDW of the form
\begin{eqnarray}\label{cdw_state}
|\Psi^{\uparrow\downarrow,0}_{\rm D}\rangle=\prod_{j=1}^{L/2}\hat{c}^{\dagger}_{2j-1,\uparrow}\hat{c}^{\dagger}_{2j-1,\downarrow}|0\rangle=|\uparrow\downarrow\ 0\ \uparrow\downarrow\ 0\cdots\uparrow\downarrow\ 0\ \rangle,
\end{eqnarray}
where $|0\rangle$ is the vacuum. This state is composed of two sublattices; the one of odd sites is fully occupied and that of even sites is empty. It corresponds to the ground state of an EFH model in the limit $V\to\infty$ with finite $U$.  

The state of~\eref{cdw_state} is degenerate with the product
\begin{eqnarray}\label{cdw_state_2}
|\Psi^{0,\uparrow\downarrow}_{\rm D}\rangle=\prod_{j=1}^{L/2}\hat{c}^{\dagger}_{2j,\uparrow}\hat{c}^{\dagger}_{2j,\downarrow}|0\rangle=|0\ \uparrow\downarrow\ 0\ \uparrow\downarrow\cdots0\ \uparrow\downarrow\rangle.
\end{eqnarray}
To discuss DQPTs arising from degenerate states, an extension to the initial description is necessary. Namely, the full return probability to the ground state manifold is~\cite{heyl2014prl,heyl2018rev} 
\begin{eqnarray}
P(t)=\mathcal{L}_{\uparrow\downarrow,0}(t)+\mathcal{L}_{0,\uparrow\downarrow}(t)
\end{eqnarray}
where $\mathcal{L}_{\eta}(t)=|\langle\eta|\psi_0(t)\rangle|^2$, and $\eta=\uparrow\downarrow,0$ and $\eta=0,\uparrow\downarrow$ denote the two degenerate ground states~\eref{cdw_state} and~\eref{cdw_state_2}, respectively. Similarly to the Loschmidt echo~\eref{echo}, each probability $\mathcal{L}_{\eta}(t)=\exp[-L\lambda_{\eta}(t)]$ is given by an intensive rate function $\lambda_{\eta}(t)$. In the thermodynamic limit, one $\mathcal{L}_{\eta}(t)$ will dominate over the other, so 
\begin{eqnarray} \label{lambda_degen}
\lambda(t)=-\lim_{L\to\infty}\frac{1}{L}\log[P(t)]=\min_{\eta}\lambda_{\eta}(t).
\end{eqnarray} 
The crossing of the two $\mathcal{L}_{\eta}(t)$, resulting in a kink, corresponds to a DQPT. This is exemplified in figure \ref{CDW_hubbard} for $U=1$ and $V=0$, where the dashed and line-dotted lines denote the rate functions associated to the two degenerate states. The first two DQPTs result from crossings of the rate functions $\lambda_{\eta}(t)$, while the third DQPT is a nonanalyticity of the dominating function $\lambda_{\uparrow\downarrow,0}(t)$.

At the crossing point, the ground state symmetry (initially broken by performing the time evolution of state~\eref{cdw_state}) is restored, which is manifested not only in the return probability $P(t)$ but also in observables; namely a vanishing order parameter at this point is expected. This foretells the importance of the dynamics of the mean imbalance $I(t)$ for the initial state \eref{cdw_state}.

Given that the initial state we consider here is uncorrelated, we could reach larger system sizes for the quench dynamics than in cases discussed in the other Sections. Thus here we report density matrix renormalization group results for $L=256$ sites. 

\subsubsection{Fermi-Hubbard limit $V=0$.} \label{dyn_cdw}

\begin{figure}[t]
\begin{center}
\includegraphics[scale=0.75]{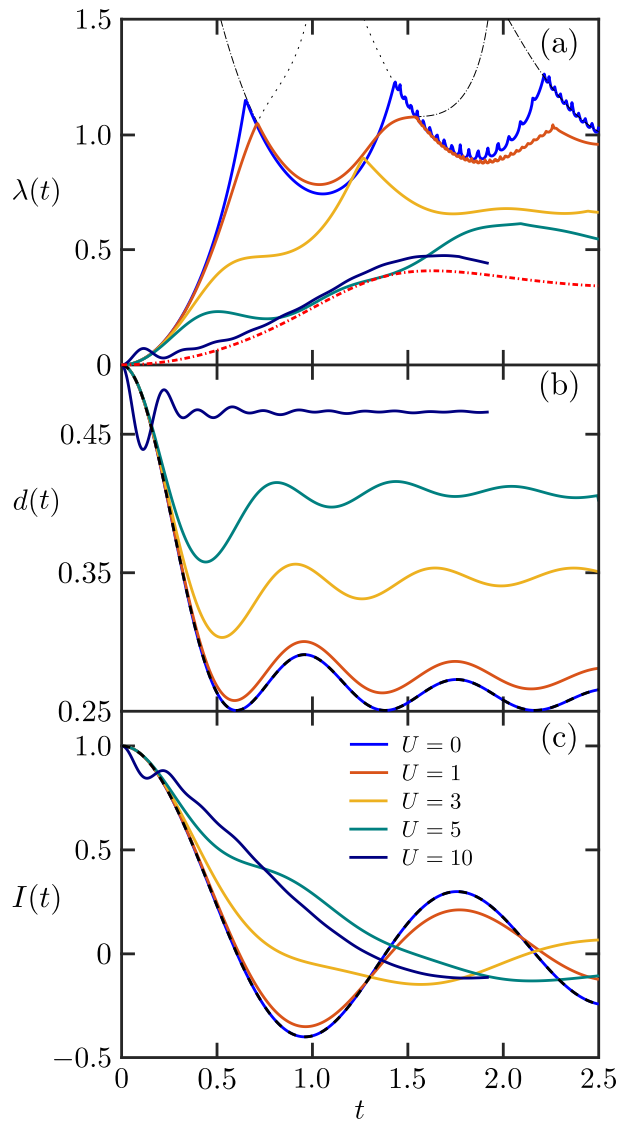}
\end{center}
\caption{(a) Rate function $\lambda(t)$ as a function of time for quenches of the Fermi-Hubbard model, with an initial product CDW state of $L=256$ sites. The thin black dashed and line-dotted lines correspond to the rate functions $\lambda_{\eta}(t)$ of the two degenerate ground states $\eta=\uparrow\downarrow,0$ and $\eta=0,\uparrow\downarrow$, respectively, for $U=1$. The thick red line-dotted curve corresponds to the rate function for a Heisenberg lattice of size $L=512$; its time scale, and that of $U=10$, is in units of $J_{{\rm ex}}^{-1}$ for better depiction. (b) Time evolution of double occupation. (c) Time evolution of population imbalance. The dashed black lines correspond to the analytical results of \eref{exact_d_I_noninteracting} for $U=0$.}
\label{CDW_hubbard}
\end{figure}

First we consider the case $V=0$ for the time-evolution Hamiltonian. The understanding of these results, shown in figure \ref{CDW_hubbard}, benefits from a key outcome of the symmetries of the Fermi-Hubbard model~\cite{enss2012njp}: the dynamics of the CDW can be directly related to that of the N\'eel state
\begin{eqnarray}\label{neel_state}
|\Psi_{\rm N}\rangle=\prod_{j=1}^{L/2}\hat{c}^{\dagger}_{2j-1,\uparrow}\hat{c}^{\dagger}_{2j,\downarrow}|0\rangle=|\uparrow\ \downarrow\ \uparrow\ \downarrow\cdots\uparrow\ \downarrow\ \rangle.
\end{eqnarray}
On the one hand, we show in \ref{appendixC} that the Loschmidt echo is equal for both states \eref{cdw_state} and \eref{neel_state} when the dynamics is performed under the same Fermi-Hubbard Hamiltonian; we have verified this with our simulations. On the other hand, it has been previously observed that assuming an even number of sites $L$, both states are related by a symmetry transformation (see \ref{appendixC}), and that the expectation values during the time evolution processes are related by~\cite{enss2012njp}
\begin{eqnarray}
d_U^{{\rm D}}(t)=d_{-U}^{{\rm D}}(t)=\frac{1}{2}-d_U^{{\rm N}}(t),\label{d_UD}\\
m_U^{{\rm N}}(t)=m_{-U}^{{\rm N}}(t)=-I_U^{{\rm D}}(t)/2,\label{symm_d_I}
\end{eqnarray}
where the indices $D$ and $N$ indicate the initial CDW and N\'eel states, respectively, and $m(t)$ is the average staggered magnetization, given by
\begin{eqnarray} \label{mean_m}
m(t)=\frac{1}{2L}\sum_{j=1}^L(-1)^j\langle\hat{n}_{j\uparrow}-\hat{n}_{j\downarrow}\rangle,
\end{eqnarray}
which for the CDW melting is zero due to spin inversion symmetry (i.e. there is no direction preference for magnetization alignment). Thus the time evolution of the expectation values of interest is independent of the sign of $U$, and the known physics of the melting of the one-dimensional N\'eel state governed by the Fermi-Hubbard model can be directly applied to that of the product CDW~\cite{enss2012njp,heidrich_meisner2015pra}.

The rate function of \eref{lambda_degen} for several values of $U$ is depicted in figure \ref{CDW_hubbard}(a); the mean double occupation and imbalance are shown in figures \ref{CDW_hubbard}(b) and (c), respectively. We start by discussing the $U=0$ case, given that analytical results for DQPTs and dynamics of observables are known. The former was studied in \cite{andraschko2014prb}, formulated in terms of the $XXZ$ model with zero anisotropy (i.e. zero interaction in the spinless-fermion picture) for an initial N\'eel state. For finite systems several small peaks emerge, corresponding to times where the initial and evolved states are orthogonal; these are seen in figure \ref{CDW_hubbard}(a) from $t\approx1.4$. However, as verified from simulations for different system sizes (see figure \ref{SizeScalingCDW} in \ref{appendixA}), these peaks decrease as the system size increases. Finally, in the thermodynamic limit only the cusps remain, indicating the DQPTs. Considering the full ground state manifold, the corresponding Fisher zeros $z=R+it$ were shown to be solutions of \cite{andraschko2014prb}
\begin{eqnarray} \label{solution_sirker}
0=\int_0^{\pi/2}dk\ln|\tan(z\epsilon_k)|^2,\quad\epsilon_{k_n}=-2J\cos(k_n+a),
\end{eqnarray}
with $a=\pi/L$, $k_n=2\pi n/L$, $n$ integer, and where the thermodynamic limit is to be taken. The real critical times $t_j^*$ correspond to the $R=0$ solutions, some of which are shown in table \ref{table_dqpt_cdw}. The three DQPTs obtained in our simulations ($t_1^*=0.650,t_2^*=1.430,t_3^*=2.215$) occur at the first three times reported in the table. In general, the values of the $U=0$ rate function shown in figure \ref{CDW_hubbard}(a) agree very well with that of \cite{andraschko2014prb} for the noninteracting $XXZ$ chain \footnote{We note that the DQPTs of \cite{andraschko2014prb} appear at twice the times of our critical times, since their hopping is half of ours. Also, our rate function is twice as large since we have the components of spin up and down orientations.}.

\begin{table}[t]
\begin{center}
\begin{tabular} {p{1.3cm} p{1cm} p{1cm} p{1cm} p{1cm} p{1cm} p{1cm} c} \hline\hline
%\begin{tabular} {p{1.2cm} c c c c c c} \hline\hline
 \centering $j$ & \centering 1 & \centering 2 & \centering 3 & \centering 4 & \centering 5 & \centering 6  &\\ \hline
 \centering $t_j^*$ & \centering 0.650 & \centering 1.429 & \centering 2.212 & \centering 2.997 & \centering 3.782 & \centering 4.566 & \\ 
 \centering $t_j^{{\rm I}}=t_j^{\rm d}$ & \centering 0.601 & \centering 1.380 & \centering 2.163 & \centering 2.948 & \centering 3.733 & \centering 4.518 & \\ 
 \centering $t_j^*-t_j^{{\rm I}}$ & \centering 0.049 & \centering 0.049 & \centering 0.049 & \centering 0.049 & \centering 0.049 & \centering 0.048 & \\ \hline\hline
\end{tabular}
\caption{Analytical results of critical times of DQPTs ($t_j^*$), times of zero imbalance ($t_j^{{\rm I}}$), and their difference, for an initial CDW state and time evolution governed by a noninteracting Fermi-Hubbard Hamiltonian.}
\label{table_dqpt_cdw}
\end{center}
\end{table}

Considering the symmetry restoration argument previously established, we now evaluate whether the times $t_j^{{\rm I}}$ of the zeros of the mean charge imbalance $I(t)$, shown in figure \ref{CDW_hubbard}(b), correspond to the critical times of the DQPTs. To answer this question, we note that the time evolution of the observables of interest can be obtained exactly in one dimension for $U=0$, and are given by~\cite{enss2012njp}
\begin{eqnarray} \label{exact_d_I_noninteracting}
d(t)=\frac{1+\mathcal{J}^2_0(4Jt)}{4},\quad I(t)=\mathcal{J}_0(4Jt),
\end{eqnarray}
with $\mathcal{J}_0$ the Bessel function of the first kind. As depicted in figures~\ref{CDW_hubbard}(b) and ~\ref{CDW_hubbard}(c), our numerical simulations agree very well with these analytical results, and the zeros of the mean imbalance agree with the times $t_j^{\rm d}$ where the mean double occupation $d(t)$ features the local minimal value $1/4$. In table \ref{table_dqpt_cdw} we report the times where these extremal expectation values take place, and see that they are not the same critical times of the DQPTs, but are shifted by a constant factor. Thus, the DQPTs and the time evolution of observables are governed by the same time scales. This is a clear example where both types of quantities are directly related.

Now we discuss the impact of turning the local interaction $U>0$ on. As in the noninteracting limit, the dynamics of the imbalance and double occupation can be extracted from those of the magnetization and double occupation for an initial N\'eel state, which have been previously studied \cite{enss2012njp,heidrich_meisner2015pra}. For finite $U/J\lesssim1$, both $d(t)$ and $I(t)$ have a similar behavior to those of the noninteracting case; the dynamics is highly-oscillating and rapidly tends to the steady-state values \cite{heidrich_meisner2015pra}. As shown in figure~\ref{CDW_hubbard}(a) for $U=1$, $\lambda(t)$ is also close to that of $U=0$. However, qualitative differences in the dynamics start to emerge. First, the zeros of the charge imbalance no longer agree with the double occupation minima (even though they remain close). More importantly, the connection between DQPTs and observables becomes less clear. Compared to the results in table \ref{table_dqpt_cdw}, the first zero of the imbalance moves to a slightly higher value ($t_1^{{\rm I}}=0.615$) while the first critical time increases more ($t_1^*=0.710$); also, the subtractions of consecutive times $\Delta t_j^*$, $\Delta t_j^{\rm d}$ and $\Delta t_j^{{\rm I}}$ do not agree anymore. When the system has stronger interactions these differences are accentuated, as already evident for $U=3$, so the connection between DQPTs and observable dynamics is entirely lost.

The dynamics of expectation values for even stronger interactions $U/J\gg1$ has also been previously analyzed. Here different types of evolutions are seen. Namely, for short times $Jt\lesssim0.5$, second-order time-dependent perturbation theory can be used to characterize the rapidly-decaying oscillations~\cite{heidrich_meisner2015pra}: 
\begin{equation}
\eqalign{d(t)=\frac{1}{2}-\frac{8J^2}{U^2}\sin^2\left(\frac{Ut}{2}\right),\qquad I(t)=1-\frac{16J^2}{U^2}\sin^2\left(\frac{Ut}{2}\right).}
\end{equation}
For long times the relaxation of the imbalance is slower than that of the double occupation, where the former is governed by the effective exchange interaction $J_{{\rm ex}}=4J^2/U$. Furthermore, $d(t)$ relaxes showing very small oscillations with frequency $U$ around its steady-state value, while $I(t)$ features wide oscillations with frequency $\propto J_{{\rm ex}}$~\cite{heidrich_meisner2015pra}. The results shown in figures~\ref{CDW_hubbard}(b) and~\ref{CDW_hubbard}(c) for $U=5,10$ agree with this description. Regarding $\lambda(t)$, depicted in figure~\ref{CDW_hubbard}(a), its first critical time $t_1^*$ increases with $U$; for example, that of $U=10$ is $t_1^*=4.42$ ($J_{{\rm ex}}t_1^*=1.77$) \footnote{This simulation was performed with $\delta t=0.01$ so long-enough times to see the crossing were reached in a reasonable computation time.}. Also, the nonanalytic behavior of the rate function, resulting from the crossing of both $\lambda_{\eta}(t)$, becomes less sharp. In addition, no coincidence with particular characteristics of $d(t)$ or $I(t)$ is found.

In the limit of very large $U/J$ the melting of the CDW state is essentially described by the Heisenberg model 
\begin{eqnarray} \label{heis_hami}
H_{{\rm Heis}}=J_{{\rm ex}}\sum_i\vec{S}_i\cdot\vec{S}_{i+1},
\end{eqnarray}
which is derived from the Fermi-Hubbard model by a Schrieffer-Wolf transformation \cite{heidrich_meisner2015pra,sw1966pr}. Thus, the double occupation of the evolving CDW is $d(t)\approx1/2$ (equivalent to impeded double occupation in the N\'eel state dynamics, see equation~\eref{d_UD}), and the long-time evolution of its charge imbalance is entirely governed by $J_{{\rm ex}}$. The rate function for a quench from a N\'eel state to the Heisenberg Hamiltonian is shown in figure~\ref{CDW_hubbard}(a), towards which results of increasing $U$ approach. Up to time $J_{{\rm ex}} t=2.5$ no DQPT is seen; in fact, at $J_{{\rm ex}}t\approx2$ the two curves of $\lambda_{\eta}(t)$ get very close but separate afterwards without crossing. Whether this eventually occurs in the Fermi-Hubbard model for very large yet finite interactions remains an open question. However, even if DQPTs still emerge, they do so for times so long that their search becomes impractical \footnote{We note that DQPTs for an initial N\'eel state and evolution governed by the Heisenberg model \eref{heis_hami} were studied in Ref. \cite{heyl2014prl}. The result was qualitatively reproduced with our simulations for sizes $L=24,64,128$, where the two curves of $\lambda_{\eta}(t)$ cross. From $L=256$ this no longer occurs, indicating that the result of Ref. \cite{heyl2014prl} for the symmetric model is an artifact of the small system considered.}.   
%Thus the DQPT for the Fermi-Hubbard model taking place at $J_{{\rm ex}}t_1^*$ and increasing values of $U$ will no longer occur in the $U\to\infty$ limit. Whether DQPTs emerge at much longer times remains an open question 

\subsubsection{Finite nearest-neighbor interaction $V\neq0$.} \label{dyn_cdw_V}

\begin{figure}[t]
\begin{center}
\includegraphics[scale=0.75]{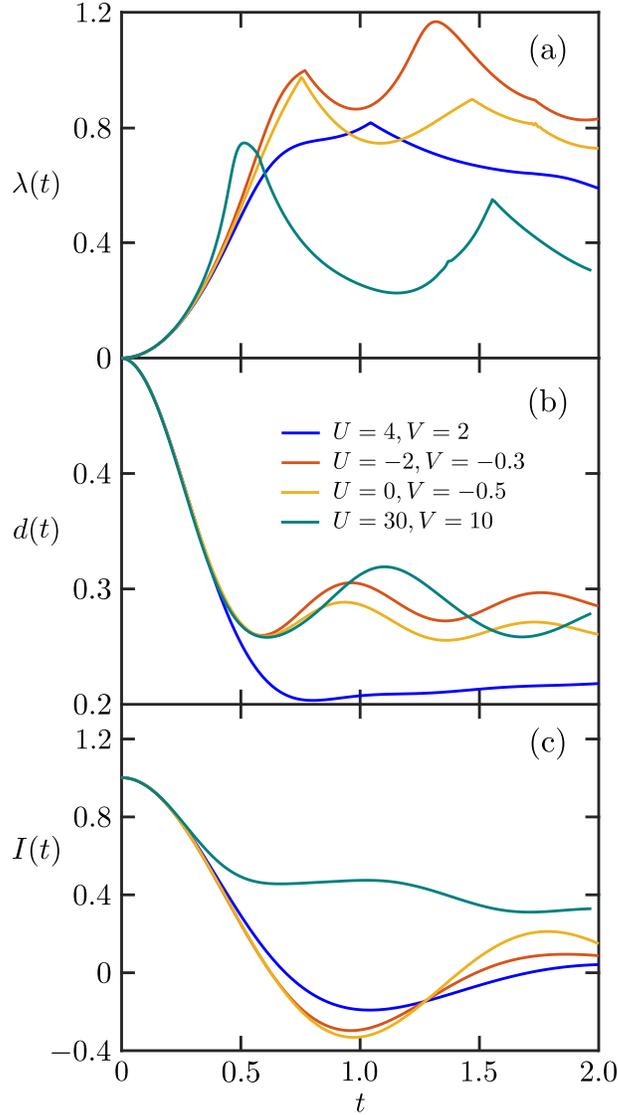}
\end{center}
\caption{(a) Rate function $\lambda(t)$ as a function of time for quenches of the extended Fermi-Hubbard model, with an initial product CDW state of $L=256$ sites. (b) Time evolution of double occupation. (c) Time evolution of population imbalance.}
\label{CDW_efh}
\end{figure}

When setting $V>0$ the dynamics becomes more involved. Furthermore, the symmetry discussed in \ref{appendixC} is no longer valid, as the unitary transformation encoding it leads to a spin-type $\hat{S}_i^z\hat{S}_{i+1}^z$ term when acting on the nearest-neighbor interaction of the Hamiltonian \cite{enss2012njp}. Thus the results for an initial N\'eel state are different to those of the product CDW. We simulated the dynamics for quenches from the product CDW to Hamiltonians corresponding to several different phases. Figure \ref{CDW_efh}(a) shows some examples of rate functions that feature DQPTs; note that these emerge on early times for cases in which no nonanalyticities were observed for the finite-coupling CDW initial state of section \ref{dqpt_finite_cdw}. Figures \ref{CDW_efh}(b) and \ref{CDW_efh}(c) present the corresponding mean double occupation and population imbalance. 

For quenches taking Hamiltonians with superconducting ground states ($U=-2$ and $V=-0.3$ for SS, $U=0$ and $V=-0.5$ for TS), the double occupation and the population imbalance are highly oscillating in time, being similar to that of the noninteracting case. On the other hand, the location of DQPTs is not evidently related to that of the zeros of $I(t)$ or the local minima of $d(t)$; the same can be said for the very-different time separation between consecutive cusps. 

For a quench Hamiltonian with stronger interactions and BOW ground state ($U=4$, $V=2$), where the oscillations of the observables are heavily damped, the DQPT is very close to the minimal imbalance (both at time $\approx1.040$). However, this coincidence is not generic of quenches to Hamiltonians with this type of ground state but accidental; indeed, different BOW quenches (e.g. to $U=2$, $V=1$), which also feature a DQPT, do not present this agreement (not shown). Finally, for an extreme SDW case ($U=30$ and $V=10$), where the charge imbalance approaches its steady state value on a larger time scale than the double occupation, the clear DQPT does not coincide with any notable behavior of the observables.

\section{DQPT induced by periodic driving} \label{floquet}
\begin{figure}[t]
\begin{center}
\includegraphics[scale=0.75]{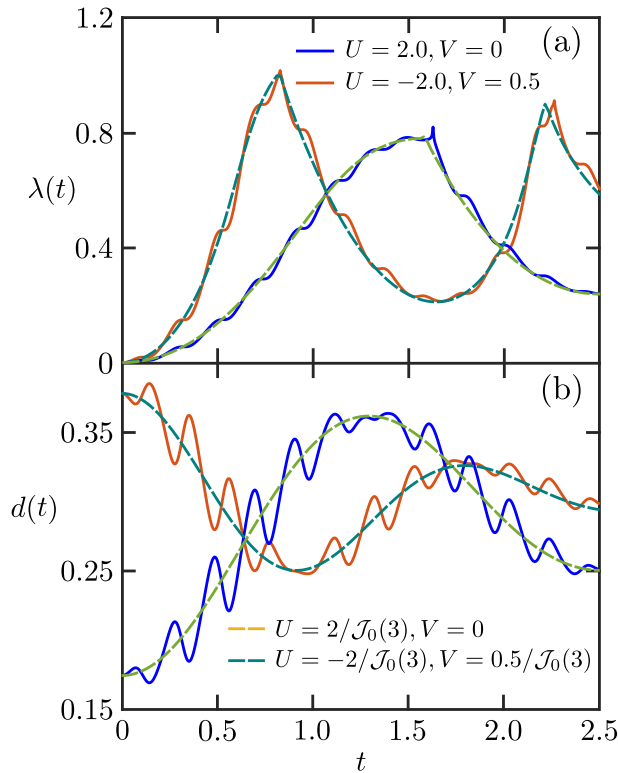}
\end{center}
\caption{DQPTs induced by periodic high-frequency driving, with $\omega=15$ and $A=3$. (a) Rate function $\lambda(t)$ as a function of time. (b) Time evolution of double occupation $d(t)$. The solid lines correspond to the results for the driven systems, and the dashed lines to those of the lattices with effective Floquet Hamiltonian.}
\label{HighFreqDriving}
\end{figure}

Finally, we discuss an alternative and far less explored method to induce DQPTs. This corresponds to driving the system by an external periodic excitation~\cite{flaschner2018nat,yang2019prb,zamani2020prb,jafari2021pra,zhou2021jpcm,hamazaki2021nat}, where the average dynamics can be described by an effective (Floquet) Hamiltonian. To perform this study, we first calculate the ground state of the EFH model at half filling for particular values of $U_0$ and $V_0$. Then we perform a real time evolution with $U=U_0$, $V=V_0$, and a time-periodic electric field of the form
\begin{eqnarray} \label{driving}
H_{driv}=E\sum_jj\sin(\omega t)\hat{n}_j,
\end{eqnarray}
with frequency $\omega$. Moving to a rotating frame, the time-periodic electric field is transformed into a time-dependent hopping parameter given by
\begin{eqnarray}
J(t)=Je^{iA\cos(\omega t)},\quad {\rm with}\quad A=E/\omega.
\end{eqnarray} 
Thus the hopping is multiplied by a Peierls phase. In this way, the emerging DQPTs are a direct consequence of the driving protocol, which is controlled by the frequency $\omega$ and the amplitude $A$.

To discuss these Floquet DQPTs, we focus on a high-frequency driving ($\omega=15$ and $A=3$), which is known to correspond to an effective Hamiltonian in which the hopping is renormalized by the Bessel function $\mathcal{J}_0(A)$~\cite{tsuji2011prl,we_dmft,kalthoff2019prb}; this is equivalent to having new density-density interactions $U/\mathcal{J}_0(A)$ and $V/\mathcal{J}_0(A)$. The results are shown in figure~\ref{HighFreqDriving} for two initial states in different phases, namely an insulating SDW state for the Fermi-Hubbard Hamiltonian and a CDW for the extended model; we see in both cases that the evolution of the double occupation and the rate function is very well captured, in average, by that of the effective Hamiltonian. 

For a SDW initial state with $U_0=2$ and $V_0=0$ the effective Floquet Hamiltonian has $U=-2/\mathcal{J}_0(3)\approx-7.69$ and $V=0$, which corresponds to a metallic ground state; considering the results of section \ref{ini_inter_hubbard}, a DQPT is anticipated, and indeed observed in figure~\ref{HighFreqDriving}(a). A similar result is obtained for a CDW case with $U_0=-2$ and $V_0=0.5$, since the effective Hamiltonian corresponds to $U=2/\mathcal{J}_0(3)\approx7.69$ and $V=0.5/\mathcal{J}_0(3)\approx-1.92$, which has a ground state deep in the SDW regime. In the light of the results of section~\ref{dqpt_finite_cdw}, DQPTs are expected and observed, as depicted in figure~\ref{HighFreqDriving}(a). 
%Similarly to several DQPTs already discussed, in neither of these cases there is a clear one-to-one correspondence between the critical times of the transitions and special features of the observable under consideration, as seen in figure~\ref{HighFreqDriving}(b).

Our results thus show that the form and time of occurrence of DQPTs can be controlled with a periodic external field, by the amplitude of the driving. In addition to the existing capabilities for implementing Fermi-Hubbard models in cold-atom architectures~\cite{schneider2012nat,ronzheimer2013prl,schreiber2015sci,scherg2018prl,tarruell2018}, recent advances in simulating periodic driving of the form~\eref{driving} by mechanical shaking~\cite{gorg2018nat,messer2018prl,sandholzer2019prl,viebahn2021prx} make our proposal a very-promising scheme for observing DQPTs in experiments. 

\section{Conclusions} \label{conclu}

The rich ground state phase diagram of systems of interacting fermions makes them an ideal testbed for observing DQPTs and establishing connections to particular characteristics of observables. With this in mind, we performed an analysis of DQPTs in the one-dimensional EFH model at half filling, for several interaction regimes.

We focused our discussion on initial noninteracting, metallic and CDW ordered states. In these regimes, DQPTs clearly emerge on two main general scenarios. The first one corresponds to initial states with zero or weak on-site interaction $U_0$ and strong final interactions, with zero (figures \ref{Rate_Hubbard_V0_Ugs0} and \ref{Rate_Hubbard_V0_Ugs-2}) or finite nearest-neighbor coupling $V$ (figure \ref{CDW_U-2V05}). The second one corresponds to initial ground states of strong interactions, namely large finite coupling $U_0$ (figure \ref{Rate_Hubbard_V0_Ugs-10}) and infinite attraction $V_0$ (product CDW, figures \ref{CDW_hubbard} and \ref{CDW_efh}), and much weaker final interactions. Thus, DQPTs are favored when the initial and final Hamiltonians involved in the quench correspond to opposed interacting regimes. 

A clear coincidence of DQPTs and particular properties of observables emerges in a much more reduced parameter space. Two main regimes were identified. The first one is that of zero or weak initial interactions and very strong final interactions (figures \ref{Rate_Hubbard_V0_Ugs0} and \ref{Rate_Hubbard_V0_Ugs-2}). Here the evolution of the rate function and the double occupation display the same timescale, given by $U$, and nonanalyticities of the former almost exactly coincide with minima of the latter. Introducing weak nearest-neighbor interactions degrades this correspondence (figure \ref{CDW_U-2V05}), but it is still observed as long as $U$ is largely dominant. The second regime corresponds to an initial product CDW (effectively equivalent to a N\'eel state) and a final noninteracting Hamiltonian (figure \ref{CDW_hubbard}), where the vanishing of the charge imbalance agrees (up to a constant shift) with the occurrence of DQPTs. This connection is broken when interactions in the final Hamiltonian are introduced. These findings would be crucial for an experimental demonstration of DQPTs and their coincidence with particular features of observables with interacting fermions.

%We initially considered the standard Fermi-Hubbard Hamiltonian ($V_0=V=0$). For a noninteracting initial state ($U_0=0$), transitions are found for strong interactions $U>0$, and agree with local minima of the double occupation. A one-to-one correspondence between both becomes less clear for a weakly-attractive initial state ($U_0=-2$), and blurs even further for a stronger-interacting case ($U_0=-10$), for which DQPTs are only seen for weak $U>0$. Notably, the same dynamics of the rate function emerges when the signs of the ground-state and time-evolution Hamiltonians are changed.

%Subsequently we analyzed DQPTs within the extended model ($V\neq0$) from initial CDWs. When starting from a state with weak nearest-neighbor repulsion ($V_0=0.5$), transitions were found for time-evolving Hamiltonians deep into the SDW and PS regimes. A connection between nonanalyticities of the rate function and special characteristics of observables seems to emerge when one interaction is largely dominant. A different correspondence is seen for some DQPTs when starting from a product CDW. This is particularly transparent for an evolution governed by a noninteracting Hamiltonian, for both the population imbalance and the mean double occupation. In addition, DQPTs are observed here even for quenches in which the first CDW did not feature them, namely to BOW, SS and TS regimes.

Finally, we demonstrated the emergence of DQPTs solely from a time-periodic on-site potential. Performing a high-frequency modulation on initial insulating SDW and CDW states, DQPTs were induced and captured by a time evolution given by an effective Floquet Hamiltonian with renormalized interactions. 

Our results pave the way for experimentally observing DQPTs in state-of-the-art quantum simulators. On the one hand, protocols for obtaining the Loschmidt echo in ultracold atom setups have been proposed \cite{daley2012prl,pichler2013njp,singh2019prx}. Also, Fermi-Hubbard systems~\cite{tarruell2018} and product CDWs~\cite{schreiber2015sci} have been implemented in fermionic optical lattices. Furthermore, existing capabilities for performing time-periodic modulation~\cite{gorg2018nat,messer2018prl,sandholzer2019prl,viebahn2021prx} makes the observation of Floquet DQPTs in such platforms very promising. On the theoretical side, our work motivates the exploration of DQPTs from different phases, e.g. superconducting or BOW states, and in richer models such as Bose-Fermi mixtures, which are experimentally available \cite{sugawa2011nat,ferrier2014science,lous2018prl} and efficiently simulable \cite{garttner2019pra,jj_berislav_2021}. Resonant driving could be also exploited, since its effective Hamiltonian, being quite different to the original \cite{bukov2016prl}, seems appealing to induce alternative Floquet DQPTs. 

\ack
The author acknowledges discussions with L. Quiroga and F. Rodr\'iguez, and is thankful to S.R. Clark for his comments on the manuscript. The author also thanks the support of Banco de la Rep\'ublica, through project No. 4.392: \textit{Control de ondas de densidad de carga y superconductividad en sistemas fermi\'onicos por medio de forzamiento peri\'odico}, and of UK's Engineering and Physical Sciences Research Council (EPSRC) under grant EP/T028424/1. This work was carried out using the computational facilities of the Advanced Computing Research Centre, University of Bristol - http://www.bristol.ac.uk/acrc/, and of Universidad de los Andes High Performance Computing (HPC) Centre.

\clearpage

\appendix

\section{Perturbative dynamics for $U_0=0$, $U\gg1$ and two sites} \label{appendix0}
To quantitatively evidence that the dynamics of the double occupancy and Loschmidt echo induced by the quench with $U_0=0$ and $U\gg1$ is governed by the same periodicity, we calculate the time evolution for a two-site system. The ground state is
\begin{eqnarray}
|\psi_0\rangle=\frac{1}{2}\left[|\uparrow\downarrow,0\rangle+|\uparrow,\downarrow\rangle-|\downarrow,\uparrow\rangle+|0,\uparrow\downarrow\rangle\right],
\end{eqnarray}
with $|\uparrow\rangle$, $|\downarrow\rangle$, $|\uparrow\downarrow\rangle$ and $|0\rangle$ the local basis states corresponding to a fermion with spin up, a fermion with spin down, a double occupancy and the empty state respectively. The time-evolved state after the quantum quench with large $U$ is obtained using strong-coupling time-dependent perturbation theory \cite{frasca1998pra}. Defining the initial and perturbing Hamiltonians as
\begin{eqnarray}\label{H0}
\hat{H}_0=-J\sum_{j=1}^{L-1}\sum_{\sigma=\uparrow,\downarrow}(\hat{c}_{j,\sigma}^{\dagger}\hat{c}_{j+1,\sigma}+{\rm H.c.}),\qquad \hat{H}_U=U\sum_{j=1}^L\hat{n}_{j\uparrow}\hat{n}_{j\downarrow},
\end{eqnarray}
the time evolved state up to second order is
\begin{eqnarray}
|\psi(t)\rangle=e^{-i\hat{H}_Ut}\left[1-i\int_0^tdt_1e^{i\hat{H}_Ut_1}H_0e^{-i\hat{H}_Ut_1}\right]|\psi_0\rangle.
\end{eqnarray}
Applying the exponential operators results in phase factors multiplying states with double occupancies. For instance,
\begin{eqnarray}
e^{-i\hat{H}_Ut}|\psi_0\rangle=\frac{1}{2}\left[e^{-iUt}|\uparrow\downarrow,0\rangle+|\uparrow,\downarrow\rangle-|\downarrow,\uparrow\rangle+e^{-iUt}|0,\uparrow\downarrow\rangle\right].
\end{eqnarray}
The time evolved state is thus given by
\begin{eqnarray}
|\psi(t)\rangle=e^{-i\hat{H}_Ut}|\psi_0\rangle+\frac{2J}{U}\left(1-e^{-iUt}\right)|\psi_0\rangle.
\end{eqnarray}
Straightforward calculations lead to the time evolution of the local double occupation,
\begin{eqnarray}
d_j(t)=\frac{1}{4}+\frac{2J}{U}\left[\frac{2J}{U}-1\right]\sin^2\left(\frac{Ut}{2}\right).
\end{eqnarray}
and of the Loschmidt echo,
\begin{eqnarray}
\mathcal{L}(t)=\cos^2\left(\frac{Ut}{2}\right)+\frac{16}{U^2}\sin^2\left(\frac{Ut}{2}\right).
\end{eqnarray}
This small system does not have DQPTs. However, it clearly illustrates that the oscillatory behavior of both the double occupation and the Loschmidt echo is governed by the same timescale, and that for very large $U$, the minima of the echo (and thus the maxima of the rate function) agree with the minima of $d(t)$. As discussed in section \ref{dqpt_fh_noint}, much larger systems signaling DQPTs also present these features. 

\section{Finite-size effects on DQPTs} \label{appendixA}

\begin{figure}[t]
\begin{center}
\includegraphics[scale=0.8]{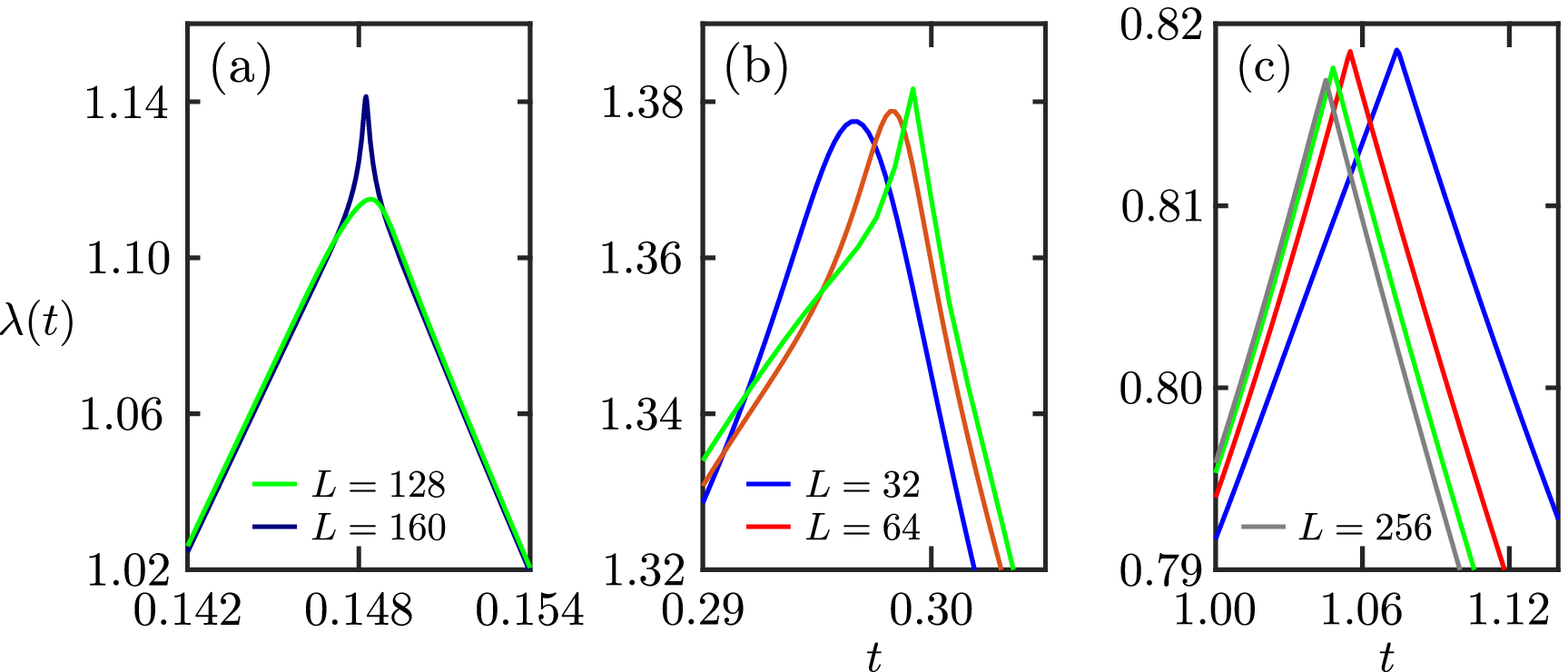}
\end{center}
\caption{Rate function $\lambda(t)$ as a function of time for different scenarios and system sizes. (a) $U_0=-2$, $U=20$, and $V_0=V=0$ (see figure \ref{Rate_Hubbard_V0_Ugs-2}). (b) $U_0=-2$, $V_0=0.5$, $U=10$ and $V=0.5$ (see figure \ref{CDW_U-2V05}). (c) Initial CDW state, $U=4$ and $V=2$ (see figure \ref{CDW_efh}). The time step was reduced up to $\delta t=2.5\times10^{-5}$ to better describe the peaks.}
\label{SizeScaling}
\end{figure}
As discussed in the main text, DQPTs are defined in the thermodynamic limit. In our simulations we have access to finite systems only; however as the system size increases, the sharp features that signal the emergence of such transitions develop, while away from these $\lambda(t)$ rapidly converges with $L$. This is shown in figure \ref{Rate_Hubbard_V0_Ugs-2}(a) for a particular quench in the Fermi-Hubbard model and $L=128,160$. The rate function is essentially identical for both system sizes during the entire time evolution except very close to the sharp peaks, as depicted in figure \ref{SizeScaling}(a) for the first one.

In figure \ref{SizeScaling} we exemplify the development of sharp features with $L$ at DQPTs for a few more cases. Panel (b) corresponds to the first peak of the rate function for an initial finite-$V$ CDW; a similar trend is observed in other cases. We also present results of different sizes for a transition emerging from the product CDW in figure \ref{SizeScaling}(c), where a sharp behavior is already seen for small systems due to the crossing of the rate functions $\lambda_{\eta}$ for both $\eta$. These results thus point towards nonanalytic features in the thermodynamic limit.

\begin{figure}[t]
\begin{center}
\includegraphics[scale=0.7]{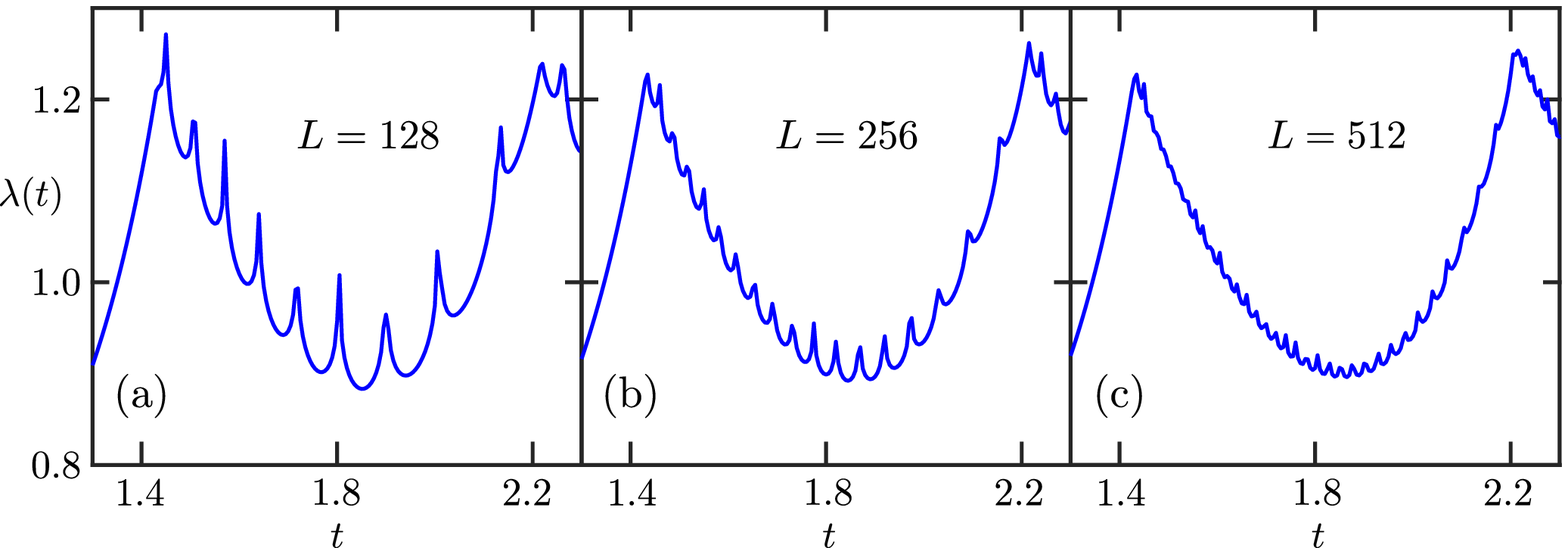}
\end{center}
\caption{Rate function $\lambda(t)$ as a function of time for an initial CDW state, $U=V=0$ and different system sizes. (a) $L=128$. (b) $L=256$. (c) $L=512$.}
\label{SizeScalingCDW}
\end{figure}

The dynamics associated to an initial product CDW and a quench to a weakly-interacting case deserves further attention, as small peaks emerge after the first DQPT due the the finite length of the system \cite{andraschko2014prb}. In figure \ref{SizeScalingCDW} we show that these peaks remain visible even for very large systems, namely $L=512$. However these are strongly suppressed as $L$ increases, pointing to their disappearance in the thermodynamic limit. 

\section{DQPT symmetry of Fermi-Hubbard model under signs exchange} \label{appendixB}
Here we demonstrate the symmetry of the rate function when the signs of the on-site density-density interactions are changed. For this we consider the ground state Hamiltonians $\hat{H}_+^{{\rm g}}$ and $\hat{H}_-^{{\rm g}}$, with couplings $+U_0$ and $-U_0$ respectively, and the time evolution Hamiltonians $\hat{H}_+^{{\rm e}}$ and $\hat{H}_-^{{\rm e}}$, with couplings $+U$ and $-U$ respectively. As discussed in \cite{enss2012njp}, the unitary transformation
\begin{eqnarray} \label{unitary}
\hat{\mathcal{U}}=\prod_j\left(\hat{c}_{j\uparrow}+(-1)^j\hat{c}_{j\uparrow}^{\dagger}\right)
\end{eqnarray}   
relates the attractive and repulsive Fermi-Hubbard models. Thus it is easily shown that
\begin{eqnarray} \label{transform_hami}
\hat{\mathcal{U}}^{\dagger}\hat{H}_+^{{\rm g}}\hat{\mathcal{U}}=\hat{H}_-^{{\rm g}},\qquad\hat{\mathcal{U}}^{\dagger}\hat{H}_+^{{\rm e}}\hat{\mathcal{U}}=\hat{H}_-^{{\rm e}},
\end{eqnarray}
and vice versa. Thus, for any time $t$,
\begin{eqnarray} \label{transform_exp_hami}
\hat{\mathcal{U}}^{\dagger}\exp(-it\hat{H}^{{\rm e}}_-)\hat{\mathcal{U}}=\exp(-it\hat{H}^{{\rm e}}_+).
\end{eqnarray}
Condition \eref{transform_hami} also leads to the relation between the ground states of the respective Hamiltonians
\begin{eqnarray} \label{transform_states}
\hat{\mathcal{U}}^{\dagger}|\psi_+^{{\rm g}}\rangle=|\psi_-^{{\rm g}}\rangle.
\end{eqnarray}
One form to show this result consists of taking the ground state as the infinite-time limit of an imaginary time evolution \cite{vidal2004prl}, namely
\begin{eqnarray}
|\psi_+^{{\rm g}}\rangle=\lim_{\tau\to\infty}\frac{\exp\left(-\tau \hat{H}_+^{{\rm g}}\right)|\phi\rangle}{||\exp\left(-\tau \hat{H}_+^{{\rm g}}\right)|\phi\rangle||},
\end{eqnarray}
for some initial state $|\phi\rangle$ which has a nonzero overlap with $|\psi_+^{{\rm g}}\rangle$. Thus we get
\begin{eqnarray}
&\hat{\mathcal{U}}^{\dagger}|\psi_+^{{\rm g}}\rangle=\lim_{\tau\to\infty}\frac{\hat{\mathcal{U}}^{\dagger}\exp\left(-\tau \hat{H}_+^{{\rm g}}\right)|\phi\rangle}{\langle\phi|\exp\left(-\tau \hat{H}_+^{{\rm g}}\right)\exp\left(-\tau \hat{H}_+^{{\rm g}}\right)|\phi\rangle^{1/2}} \nonumber\\
&=\lim_{\tau\to\infty}\frac{\exp\left(-\tau \hat{H}_-^{{\rm g}}\right)\hat{\mathcal{U}}^{\dagger}|\phi\rangle}{\langle\phi|\hat{\mathcal{U}}\exp\left(-\tau \hat{H}_-^{{\rm g}}\right)\exp\left(-\tau \hat{H}_-^{{\rm g}}\right)\hat{\mathcal{U}}^{\dagger}|\phi\rangle^{1/2}} \nonumber\\
&=\lim_{\tau\to\infty}\frac{\exp\left(-\tau \hat{H}_-^{{\rm g}}\right)\hat{\mathcal{U}}^{\dagger}|\phi\rangle}{||\exp\left(-\tau \hat{H}_-^{{\rm g}}\right)\hat{\mathcal{U}}^{\dagger}|\phi\rangle||}=|\psi_-^{{\rm g}}\rangle,
\end{eqnarray}
provided $\hat{\mathcal{U}}^{\dagger}|\phi\rangle$ has a nonzero overlap with $|\psi_-^{{\rm g}}\rangle$. Using Eqs. \eref{transform_exp_hami} and \eref{transform_states} we have, for the Loschmidt amplitude of the dynamics with initial state $|\psi_+^{{\rm g}}\rangle$ and time evolution under $\hat{H}_-^{{\rm e}}$
\begin{eqnarray} \label{evolution_Loschmidt}
\langle\psi_+^{{\rm g}}|\exp(-it\hat{H}_-^{{\rm e}})|\psi_+^{{\rm g}}\rangle&=\langle\psi_+^{{\rm g}}|\hat{\mathcal{U}}\hat{\mathcal{U}}^{\dagger}\exp(-it\hat{H}_-^{{\rm e}})\hat{\mathcal{U}}\hat{\mathcal{U}}^{\dagger}|\psi_+^{{\rm g}}\rangle\nonumber\\
&=\langle\psi_-^{{\rm g}}|\exp(-it\hat{H}_+^{{\rm e}})|\psi_-^{{\rm g}}\rangle.
\end{eqnarray}
The Loschmidt amplitude, and thus the echo and the rate function, are invariant under the change of signs of the on-site interactions of the ground state and time evolution Hamiltonians.

Similarly, the time evolution of the double occupation $\hat{d}_j=\hat{n}_{j\uparrow}\hat{n}_{j\downarrow}$ of site $j$ under the same scheme as in \eref{evolution_Loschmidt}, which we name $d_j^{+-}(t)$ (first sign in the exponent indicating $+U$ for ground state, and second sign indicating $-U$ for time evolution), is given by
\begin{eqnarray} \label{evolution_dbl}
d_j^{+-}(t)&=\langle\psi_+^{{\rm g}}|\exp(it\hat{H}_-^{{\rm e}})\hat{d}_j\exp(-it\hat{H}_-^{{\rm e}})|\psi_+^{{\rm g}}\rangle\nonumber\\
&=\langle\psi_+^{{\rm g}}|\hat{\mathcal{U}}\hat{\mathcal{U}}^{\dagger}\exp(it\hat{H}_-^{{\rm e}})\hat{\mathcal{U}}\hat{\mathcal{U}}^{\dagger}\hat{d}_j\hat{\mathcal{U}}\hat{\mathcal{U}}^{\dagger}\exp(-it\hat{H}_-^{{\rm e}})\hat{\mathcal{U}}\hat{\mathcal{U}}^{\dagger}|\psi_+^{{\rm g}}\rangle\nonumber\\
&=\langle\psi_-^{{\rm g}}|\exp(it\hat{H}_+^{{\rm e}})(\hat{n}_{j\downarrow}-\hat{d}_j)\exp(-it\hat{H}_+^{{\rm e}})|\psi_-^{{\rm g}}\rangle,
\end{eqnarray}
where in the last equality we used that $\hat{\mathcal{U}}^{\dagger}\hat{d}_j\hat{\mathcal{U}}=\hat{n}_{j\downarrow}-\hat{d}_j$. Performing an average of the terms of \eref{evolution_dbl} over all sites $j$, and considering that the number on fermions with spin down is a constant of motion with value given by \eref{symm}, we finally get
\begin{eqnarray}
d_j^{+-}(t)\equiv\frac{1}{L}\sum_{j=1}^Ld_j^{+-}(t)=\frac{1}{2}-\sum_{j=1}^Ld_j^{-+}(t)\equiv \frac{1}{2}-d_j^{+-}(t).
\end{eqnarray}
Thus the average double occupations for the evolution schemes with exchanged interaction signs are symmetric with respect to $1/4$.

\section{Symmetry between dynamics of CDW and N\'eel states} \label{appendixC}
In \cite{enss2012njp} it was shown that the product CDW and N\'eel states are related through the unitary \eref{unitary} by 
\begin{eqnarray}
|\Psi_{\rm N}\rangle=\hat{\mathcal{U}}^{\dagger}|\Psi_{\rm D}\rangle,\qquad|\Psi_{\rm D}\rangle=\hat{\mathcal{U}}^{\dagger}|\Psi_{\rm N}\rangle.
\end{eqnarray}
Furthermore, based on the unitary transformation and on the time reversal invariance of the expectation values, Eqs. \eref{symm_d_I} and \eref{d_UD} were demonstrated. Now we show that the Loschmidt echoes, and thus the rate function, is equal when a quench is performed on both states with the same Fermi-Hubbard Hamiltonian (the Loschmidt amplitudes themselves might be different, as observed from our simulations). Starting from the echo for a N\'eel state:
\begin{eqnarray}
&|\langle\Psi_{\rm N}|\exp(-it\hat{H}^{{\rm e}})|\Psi_{\rm N}\rangle|^2=\langle\Psi_{\rm N}|\exp(-it\hat{H}^{{\rm e}})|\Psi_{\rm N}\rangle\langle\Psi_{\rm N}|\exp(it\hat{H}^{{\rm e}})|\Psi_{\rm N}\rangle\nonumber\\
&=\langle\Psi_{\rm D}|\hat{\mathcal{U}}\exp(-it\hat{H}^{{\rm e}})\hat{\mathcal{U}}^{\dagger}|\Psi_{\rm D}\rangle\langle\Psi_{\rm D}|\hat{\mathcal{U}}\exp(it\hat{H}^{{\rm e}})\hat{\mathcal{U}}^{\dagger}|\Psi_{\rm D}\rangle\nonumber\\
&=|\langle\Psi_{\rm D}|\exp(-it\hat{H}^{{\rm e}})|\Psi_{\rm D}\rangle|^2,
\end{eqnarray}
thus obtaining the echo for the product CDW.\\

\bibliographystyle{unsrtnt}

\bibliography{EFH-Dynamics_Bib}

\end{document}